\documentclass[12pt,preprint]{emulateapj}
\begin{document}
\title{\lowercase{$z\sim7-10$} Galaxies behind Lensing Clusters: Contrast with Field Search Results}
\author{Rychard J. Bouwens$^{2}$, Garth D. Illingworth$^{2}$, Larry
D. Bradley$^{3}$, Holland Ford$^{3}$, Marijn Franx$^{4}$, Wei
Zheng$^{3}$, Tom Broadhurst$^{5}$, Dan Coe$^{6}$, M. James Jee$^{7}$}

\affil{1 Based on observations made with the NASA/ESA Hubble Space
  Telescope, which is operated by the Association of Universities for
  Research in Astronomy, Inc., under NASA contract NAS 5-26555.  These
  observations are associated with programs \#5352, 5935, 6488, 8249,
  8882, 9289, 9452, 9717, 10150, 10154, 10200, 10325, 10504, 10863,
  10996.}
\affil{2 Astronomy Department, University of California, Santa Cruz,
CA 95064} \affil{3 Department of Physics and Astronomy, Johns Hopkins
University, 3400 North Charles Street, Baltimore, MD 21218} 
\affil{4 Sterrewacht Leiden, Leiden University, NL-2300 RA Leiden, Netherlands}

\affil{5 School of Physics and Astronomy, Tel Aviv University, Tel
Aviv 69978, Israel} \affil{6 Jet Propulsion Laboratory, California
Institute of Technology, MS 169-327, Pasadena, CA 91109} \affil{7
Department of Physics, 1 Shields Avenue, University of California,
Davis, CA 95616}

\begin{abstract}
We conduct a search for $z\gtrsim7$ dropout galaxies behind 11 massive
lensing clusters using 21 arcmin$^2$ of deep HST NICMOS, ACS, and
WFPC2 image data. In total, over this entire area, we find only one
robust $z\sim7$ $z$-dropout candidate (previously reported around
Abell 1689).  Four less robust $z$-dropout and $J$-dropout candidates
are also found.  The nature of the four weaker candidates could not be
precisely determined due to the limited depth of the available optical
data, but detailed simulations suggest that all four could be
low-redshift interlopers.  We compare these numbers with what we might
expect using the $z\sim7$ UV luminosity function (LF) determined from
field searches.  We predict 2.7 $z\sim7$ $z$-dropouts and 0.3 $z\sim9$
$J$-dropouts over our cluster search area, in reasonable agreement
with our observational results, given the small numbers.  The number
of $z\gtrsim7$ candidates we find in the present search are much lower
than has been reported in several previous studies of the prevalence
of $z\gtrsim7$ galaxies behind lensing clusters.  To understand these
differences, we examined $z\gtrsim7$ candidates in other studies and
conclude that only a small fraction are likely to be $z\gtrsim7$
galaxies.  Our findings support models that show that gravitational
lensing from clusters is of the most value for detecting galaxies at
magnitudes brighter than L$^*$ ($H\lesssim 27$) where the LF is
expected to be very steep.  Use of these clusters to constrain the
faint-end slope or determine the full LF is likely of less value due
to the shallower effective slope measured for the LF at fainter
magnitudes, as well as significant uncertainties introduced from
modelling both the gravitational lensing and incompleteness.
\end{abstract}

\section{Introduction}

Because of the great distances and extreme faintness of galaxies at
$z\gtrsim7$, as well as the high sky backgrounds, detection of
galaxies at such high redshifts remains extremely challenging. It is
not surprising that the number of robust $z$$\gtrsim$7 candidates is
still very small (see, e.g., Bouwens \& Illingworth 2006; Bouwens et
al.\ 2008; Oesch et al.\ 2008).  Gravitational lensing by galaxy
clusters has been highlighted as an efficient way of improving this
situation, due to the significant areas on the sky behind these
clusters with sizeable magnification factors to amplify light from
faint sources.  However, this advantage is offset by the greatly
reduced source plane volume in the highly magnified regions.

Because of the trade-off between depth and area, the utility of
clusters for these searches depends strongly on the slope of the
luminosity function (e.g., Broadhurst et al.\ 1995).  If the effective
slope of the LF ($-d(\log d\Phi)/d\log L$) is greater than 1, the gain
in depth more than compensates for the loss in area, increasing the
overall number of sources (e.g., Broadhurst et al.\ 1995) over that
found in the field.  Such steep slopes are found at magnitudes
brightward of $L^*$ (corresponding to H$\lesssim$27 AB mag for
$z\sim7$ galaxies).  By contrast, at fainter magnitudes (H$\gtrsim$27
for $z\sim7$ galaxies) the effective slope of the LF is not quite so
steep (e.g., the faint-end slope for the Bouwens et al.\ 2007 $z\sim6$
LF corresponds to $-d(\log \Phi)/d(\log L)\sim0.7<1$).  This trade-off
between depth and area is such that the surface density of dropouts is
lower behind clusters than in the field at faint magnitudes.  Overall
these considerations suggest that the most significant advantages will
be achieved at bright magnitudes ($>$L$^*$) where the LF is very
steep.  Shallow searches over many clusters, in particular, would seem
to be the most rewarding.

\begin{figure*}
\epsscale{1.18}
\plotone{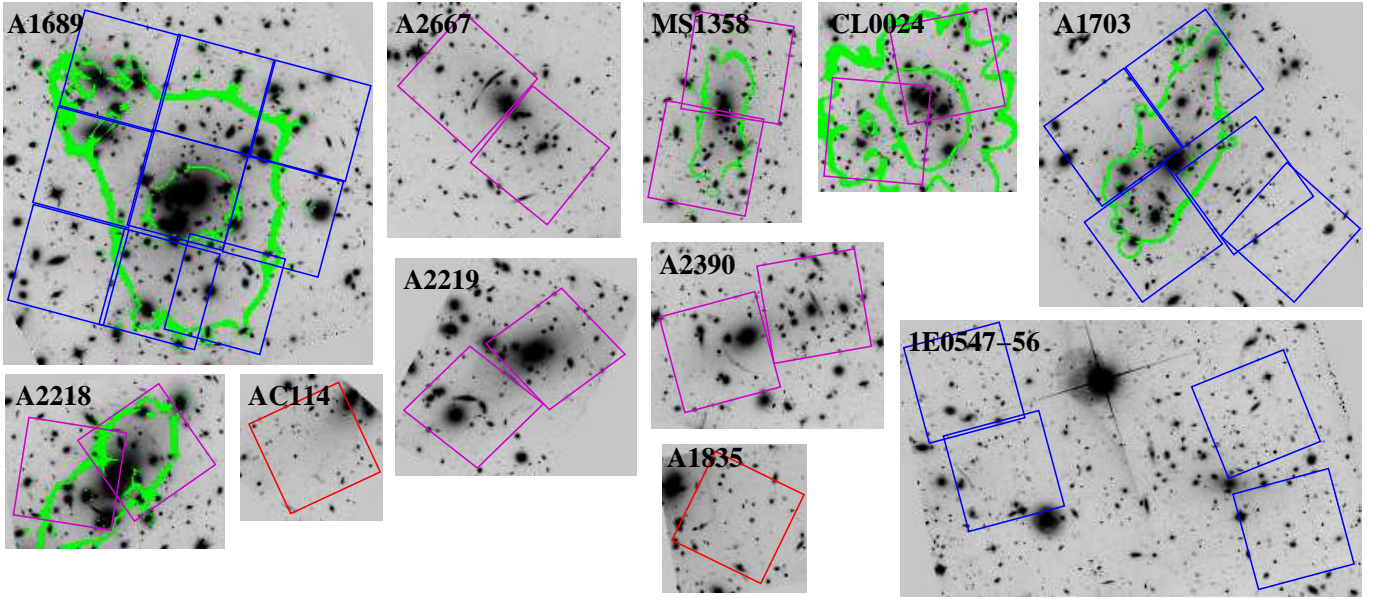}
\caption{Optical ACS images of the 11 massive lensing clusters we used
  to search for star-forming galaxies at $z\gtrsim7$.  The name of
  each cluster is included in the upper left corner of the
  corresponding cluster image.  Overlaid on these images are outlines
  of our deep NIC3 search fields, where magenta indicates that the
  field has deep ($\sim$3-5 orbit) data in both the $J_{110}$ and
  $H_{160}$ bands, blue indicates the field only has shallow
  ($\sim$1-2 orbit) data in the $J_{110}$ band, and red indicates the
  field only has deep ($\sim$4 orbit) data in the $H_{160}$ band (see
  Table~\ref{tab:obsdata}).  The green contours show the position of
  the critical curves at $z$$\sim$7 (i.e., where the magnification
  $\mu$$>$100) as determined from several published lensing models
  (Limousin et al.\ 2007; Limousin et al.\ 2008;
  El{\'{\i}}asd{\'o}ttir et al.\ 2007; Jee et al.\ 2007; Franx et
  al.\ 1997).\label{fig:obsdata}}
\end{figure*}

\begin{deluxetable*}{cccccccc}[b]
\tablecolumns{8}
\tablecaption{HST NICMOS data around massive galaxy clusters used to search for $z\gtrsim7$ galaxies.\label{tab:obsdata}}
\tablehead{
\colhead{} & \colhead{Area} & \multicolumn{4}{c}{5$\sigma$ Depth\tablenotemark{a}} & \colhead{NICMOS} & \colhead{} \\
\colhead{Name} & \colhead{(arcmin$^2$)} & \colhead{Optical\tablenotemark{b}} & \colhead{$z_{850}$} & \colhead{$J_{110}$} & \colhead{$H_{160}$} & \colhead{orbits} & \colhead{Ref\tablenotemark{c}}}
\startdata
MS1358 & 1.4 & 28.0 & 27.5 & 26.8 & 26.7 & 18 & [1] \\
CL0024 & 1.4 & 27.8 & 27.4 & 26.8 & 26.7 & 18 & [1] \\
Abell 2218 & 1.2 & 27.9 & 27.6 & 26.8 & 26.7 & 17 & [1] \\
Abell 2219 & 1.4 & 27.5 & 27.2 & 26.8 & 26.7 & 18 & [1] \\
Abell 2390 & 1.5 & 26.6\tablenotemark{d} & 26.9 & 26.8 & 26.7 & 18 & [1] \\
Abell 2667 & 1.5 & 26.5\tablenotemark{d} & 26.9 & 26.8 & 26.7 & 18 & [1] \\
Abell 1689 & 5.7 & 28.1 & 26.8 & 26.4 & --- & 19 & [2] \\
Abell 1703 & 3.1 & 27.9 & 26.7 & 26.0 & --- & 6 & --- \\
1E0657-56\tablenotemark{e} & 2.7 & 27.6 & 27.2 & 26.0 & --- & 4 & --- \\
Abell 1835 & 0.7 & 27.1\tablenotemark{d} & 27.2 & --- & 26.7 & 4 & --- \\
AC114 & 0.7 & 27.1\tablenotemark{d} & 27.2 & --- & 26.7 & 4 & --- \\
\hline
Total & 21.3 & & & & & 144 &
\enddata
\tablenotetext{a}{$5\sigma$ limits assume a $0.3''$-diameter aperture for ACS/WFPC2 and $0.6''$-diameter aperture for NICMOS.}
\tablenotetext{b}{The depth of the deepest single-band optical ($\leq0.8\mu$)
  image available over the cluster.  Note that some of the
  clusters listed here have very deep data in $\geq2$ bands
  (particularly MS1358, CL0024, Abell 2218, Abell 1689, Abell
  1703), so the effective depth of the combined optical data is often $>$0.4
  mag deeper than tabulated here.}
\tablenotetext{c}{References: [1] Richard et al. 2008,
[2] Bradley et al.\ 2008.}
\tablenotetext{d}{The optical data available over these four clusters
  are from WFPC2 and are only moderately deep.  As a result, we might
  expect a small number of low redshift interlopers to make it into
  $z\gtrsim7$ selections (\S4) conducted over each of these clusters.
  Not surprisingly, a substantial fraction of the weaker $z\gtrsim7$
  candidates in our own selection, and that of Richard et al.\ (2008)
  are found over these four clusters.}
\tablenotetext{e}{The ``Bullet'' cluster}
\end{deluxetable*}

Here we assess the promise of clusters for studying $z\gtrsim7$
galaxies by conducting a careful search for high redshift galaxies in
all the currently available HST NICMOS imaging data over 11 massive
low-redshift galaxy clusters.  Bradley et al.\ (2008) have already
conducted such a search around Abell 1689 and reported one highly
robust $z\gtrsim7$ galaxy.  Richard et al.\ (2006) examined 2 clusters
and reported 13 $z\gtrsim6$ candidates, while Richard et al.\ (2008)
have examined 6 clusters and reported 12 $z\gtrsim6$ candidates.  The
present paper represents an independent assessment of the prevalence
of these sources behind massive low-redshift clusters.  We take
advantage of $\sim$21 arcmin$^2$ of very deep NICMOS data behind 11
lensing clusters with optical ACS+WFPC2 coverage, 7 of which were
already considered in the Bradley et al.\ (2008) and Richard et
al.\ (2008) papers.  For our $z\gtrsim7$ search, we will utilize many of
the same photometric techniques we have employed over the past few
years to identify large samples of $z$$\sim$4-8 $BViz$ dropouts in the
field (Bouwens et al.\ 2006; Bouwens et al.\ 2007; Bouwens et
al.\ 2004; Bouwens et al.\ 2008).

We begin this paper with a summary of the observational data available
to search for $z\gtrsim7$ galaxies behind galaxy clusters (\S2).  In
\S3, we describe our techniques for constructing source catalogs and
doing the photometry (\S3).  In \S4, we summarize the results of our
selection and discuss sources of contamination.  In \S5, we compare
our results with (1) what we would expect based upon the $z\sim7$ LF
derived in the field and (2) the search results from other teams.  We
then conclude by discussing the implications of this study (\S6).  In
the Appendix, we provide an assessment of the $z\gtrsim7$ candidates
reported by Richard et al.\ (2008).  In particular, we assess the
merits of searching for $z\gtrsim7$ galaxies behind clusters versus
searching for these galaxies in the field (\S6).  We assume $\Omega_0
= 0.3$, $\Omega_{\Lambda} = 0.7$, $H_0 = 70\,\textrm{km/s/Mpc}$
throughout.  Although these parameters are slightly different from
those determined from the WMAP five-year results (Dunkley et
al.\ 2008), they allow for convenient comparison with other recent
results expressed in a similar manner.  The HST filters F555W, F606W,
F625W, F702W, F775W, F814W, F850LP, F110W, and F160W will be denoted
as $V_{555}$, $V_{606}$, $r_{625}$, $R_{702}$, $i_{775}$, $I_{814}$,
$z_{850}$, $J_{110}$, and $H_{160}$, respectively.  We will express
all magnitudes in the AB system (Oke \& Gunn 1983).

\section{Observational Data}

We conduct searches for dropout galaxies over the 11 low-redshift
galaxy clusters with deep near-IR NICMOS and optical ACS coverage
(Table~\ref{tab:obsdata}).  The near-IR data here were taken with the
goal of finding $z\gtrsim7$ galaxies.  The NICMOS coverage of the
first six clusters considered here (CL0024, MS1358, Abell 2218, Abell
2219, Abell 2390, Abell 2667) extends over $8.4$ arcmin$^2$ (12 NIC3
pointings) and includes very deep imaging in both the $J_{110}$ and
$H_{160}$ bands (Richard et al.\ 2008).  The NICMOS coverage of three
other clusters considered here (i.e., Abell 1689, Abell 1703, and
1E0657-56) is somewhat shallower in general, extends over $\sim$10
arcmin$^2$ (17 NIC3 pointings), and is mainly in the $J_{110}$ band.
The final two clusters considered here (Abell 1835 and AC114) only
have deep near-IR NICMOS data in the $H_{160}$ band and over one
$\sim$0.8 arcmin$^2$ NIC3 pointing per cluster.  The layout of our
NICMOS search fields is illustrated graphically in
Figure~\ref{fig:obsdata}.

Deep optical imaging data are necessary for the selection of
$z\gtrsim7$ galaxies and can be a significant limitation, if not
available.  MS1358, CL0024, Abell 1689, Abell 1703, and Abell 2218 all
possess useful optical data, with $\geq2$ orbits in each of the ACS
$g_{475}$, $r_{625}$, and $i_{775}$ bands and $\geq6$ orbits in the
ACS $z_{850}$ band.  The optical coverage of Abell 2219, Abell 2390,
Abell 2667, 1E0657-56, Abell 1835, and AC114 is generally shallower in
depth and primarily with WFPC2, though $\sim$4 orbits of ACS $z$-band
coverage are available for each.

All the available ACS and NICMOS data over these clusters are
processed into image mosaics using the ACS GTO pipeline ``apsis''
(Blakeslee et al.\ 2003) and NICMOS pipeline ``nicred.py'' (Magee et
al.\ 2007).  Reductions of the WFPC2 data are obtained from the
Canadian Astronomy Data Centre.  All reductions are registered onto
the same frame as the NICMOS data.

\section{Sample Construction}

\textit{(a) Catalog Generation:} Our procedure for generating source
catalogs and doing photometry is identical to that performed in
Bouwens et al.\ (2008).  Briefly, we run SExtractor (Bertin \& Arnouts
1996) in double-image mode to do object detection and photometry.  For
the detection image, we use the square root of the $\chi^2$ image
(Szalay et al.\ 1999), which we construct from the NICMOS $J_{110}$
and $H_{160}$ images for our $z_{850}$ dropout selection and the
NICMOS $H_{160}$ band image for our $J_{110}$ dropout selection.  Our
photometry is then conducted on our ACS, WFPC2, and NICMOS images,
which are point-spread function matched to the NICMOS $H_{160}$ image.
Colors are measured in small-scalable (Kron 1980) apertures, assuming
a Kron (1980) factor of 1.2.  These fluxes are then corrected to total
magnitudes using the light within a larger Kron (1980) aperture
(adopting a Kron factor of 2.5).  These latter corrections are made
using the square root of the $\chi^2$ image to improve the S/N.
Figure 5 of Coe et al.\ (2006) provides a graphical description of a
similar multi-stage procedure for measuring colors and total
magnitudes.  Typical aperture diameters are 0.3$''$ and 0.6$''$ for
color and total magnitude measurements, respectively.

\textit{(b) Selection Criteria:} We use the same selection criteria
for identifying star-forming galaxies at $z\gtrsim7$ that we used in
our previous work on the identification of such galaxies in field data
sets like GOODS or the HUDF (e.g., Bouwens et al.\ 2008).
Specifically, we require a $z$-dropout candidate to satisfy the
criterion $((z_{850}-J_{110})_{AB})>0.8)\wedge
((z_{850}-J_{110})_{AB}>0.8+0.4(J_{110}-H_{160})_{AB})$ where $\wedge$
represents the logical \textbf{AND} symbol.  $J$-dropout candidates
are expected to satisfy the criterion $(J_{110}-H_{160})_{AB}>1.3$.
In cases of a non-detection in the dropout band, we set the flux in
the dropout band to be equal to the $1\sigma$ upper limit.  We require
each candidate to be completely undetected ($<2\sigma$) in all
passbands blueward of the dropout band.  We also demand that each
candidate be detected at $5.5\sigma$ in the $H_{160}$ band in a
$0.5''$-diameter aperture to eliminate spurious sources.

Our selection criteria are modified slightly for clusters that do not
have NICMOS imaging in both the $J$ and $H$ bands.  For clusters with
only $J$ band imaging, we apply a $(z_{850}-J_{110})_{AB}>1.0$
criterion to select $z$$\sim$7 $z$-dropouts and for clusters with only
$H$ band imaging, we apply a $(z_{850}-H_{160})_{AB}>1.2$ criterion to
identify possible star-forming galaxies at $z\gtrsim7$.  Both criteria
should be successful in identifying candidate $z\gtrsim7$ galaxies
(albeit with a higher contamination level than selections relying on
both $J$ and $H$ data).

\begin{figure}
\epsscale{1.15} \plotone{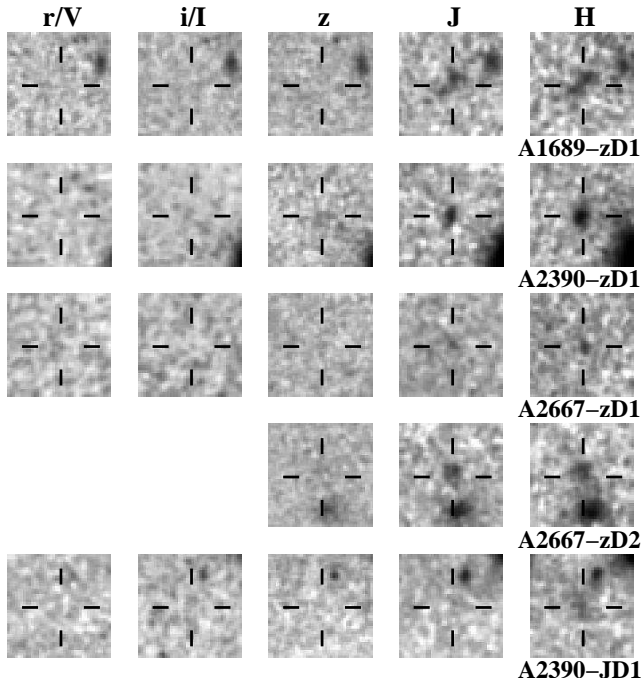}
\caption{$r_{625}$,$i_{775}$,$z_{850}$,$J_{110}$,$H_{160}$ images
  ($4'' \times 4''$) of 4 $z$$\sim$7 $z$-dropout candidates and
  one $z$$\sim$9 $J$-dropout candidate (A2390-JD1), identified behind
  11 massive lensing clusters.  We consider the top source (A1689-zD1:
  Bradley et al.\ 2008) to be the only highly robust $z\gtrsim7$
  candidate in our cluster search fields.  The leftmost two images
  correspond to the WFPC2 $V_{555}$ and $I_{814}$ bands for Abell 2390
  and the WFPC2 $V_{606}$ and $I_{814}$ bands for Abell 2667.  Deep
  WFPC2 data are not available for A2667-zD2.  \label{fig:stamp}}
\end{figure}

\begin{deluxetable*}{cccccccc}
\tablecolumns{8} \tablewidth{6.1in} \tablecaption{$z\gtrsim7$
  $z,J$-dropout candidates.\tablenotemark{*}\label{tab:candlist}} \tablehead{
  \colhead{Object ID} & \colhead{R.A.} & \colhead{Dec} &
  \colhead{$0.6\mu-J_{110}$\tablenotemark{a}} &
  \colhead{$0.8\mu-J_{110}$\tablenotemark{b}} &
  \colhead{$z_{850}-J_{110}$} & \colhead{$J_{110}-H_{160}$} &
  \colhead{$H_{160}$}} 
\startdata 
\multicolumn{8}{c}{$z\sim7$  $z$-dropouts} \\ 
A1689-zD1\tablenotemark{c} & 13:11:29.73 &
$-$01:19:20.9 & $>$2.5\tablenotemark{d} & $>$2.5\tablenotemark{d} &
$>$2.2\tablenotemark{d} & 0.6$\pm$0.2 & 24.6$\pm$0.1 \\ 
A2390-zD1\tablenotemark{e} & 21:53:34.09 & 17:41:41.1 &
$>$1.7\tablenotemark{d} & $>$1.5\tablenotemark{d} & 1.1$\pm$0.8 &
0.8$\pm$0.2 & 25.2$\pm0.2$ \\
A2667-zD1\tablenotemark{e} & 23:51:40.06
& $-$26:05:13.8 & $>$0.9\tablenotemark{d} & $>$0.4\tablenotemark{d} &
$>$1.2\tablenotemark{d} & 0.0$\pm$0.3 & 26.1$\pm$0.2 \\ 
A2667-zD2\tablenotemark{e} & 23:51:36.85 & $-$26:05:21.4 & --- &
--- & 0.9$\pm$0.4 & 0.0$\pm$0.3 & 25.6$\pm$0.2 \\ 
\multicolumn{8}{c}{$z\sim9$ $J$-dropouts} \\ 
A2390-JD1\tablenotemark{e,g} & 21:53:34.12 & 17:41:44.2 &
$>$1.2\tablenotemark{d,f} & $>$1.1\tablenotemark{d,f} &
$>$1.3\tablenotemark{d,f} & $>$1.9 & 26.0$\pm$0.2\tablenotemark{g}
\enddata 
\tablenotetext{*}{For uniformity of our analysis, all of the $z\gtrsim7$
  candidates here are from our own $z\gtrsim7$ $z$ and $J$ dropout
  searches in cluster fields and do not include independent search
  results.}

\tablenotetext{a}{This colour corresponds to
  $r_{625}-J_{110}$ for Abell 1689, $V_{555}-J_{110}$ for Abell 2390,
  and $V_{606}-J_{110}$ for Abell 2667.}  \tablenotetext{b}{This
  colour corresponds to $i_{775}-J_{110}$ for Abell 1689 and
  $I_{814}-J_{110}$ for Abell 2390 and Abell 2667.}
\tablenotetext{c}{A1689-zD1 is our most robust $z\gtrsim7$ candidate
  and was previously presented in Bradley et al.\ (2008).}
\tablenotetext{d}{Lower limits on the measured colors are
  $1\sigma$ limits.}  \tablenotetext{e}{No deep optical data are
  available for the four candidates: A2390-zD1, A2667-zD1, A2667-zD2,
  and A2390-JD1 (see Table~\ref{tab:obsdata}).  We consider all four
  to be relatively weak $z\gtrsim7$ $z$-dropout candidates.}
\tablenotetext{f}{The colours tabulated here are relative to the $H$-band, not the $J_{110}$-band, and hence are $0.6\mu-H_{160}$, $0.8\mu-H_{160}$, and $z_{850}-H_{160}$.}
\tablenotetext{g}{While this source is formally a $5.5\sigma$
  detection in our reductions (and using our photometric procedures),
there is some chance it may still be spurious.}
\end{deluxetable*}

\section{Results}

After careful application of our selection criteria to all 11 clusters
under study here, we identify 4 $z$-dropout candidates and 1
$J$-dropout candidate.  We also uncover a small number of candidates
that appeared to be promising $z\gtrsim7$ candidates (e.g., the
candidates at 00:26:37.90, 17:09:10.4 or 00:26:35.11, 17:10:10.3
behind CL0024), but which show modest ($\sim$2$\sigma$) detections in
passbands blueward of $z_{850}$ and therefore were excluded.

Postage stamps of the four candidate star-forming galaxies at
$z\gtrsim7$ are provided in Figure~\ref{fig:stamp}.  Other properties
are given in Table~\ref{tab:candlist}.  Our candidate in Abell 1689
has already been reported upon before by Bradley et al. (2008) and is
by far the strongest candidate.  The SED derived from the very deep
optical, NICMOS, and IRAC data can only be successfully fit by the SED
of a star-forming galaxy at $z\sim7.6$ (Bradley et al.\ 2008).

\textit{Possible Contamination of our Selection by Lower-Redshift
  Galaxies:} The four new sources in our selection are much less
robust $z\gtrsim7$ candidates.  All are in clusters for which the
optical data is much shallower. The $V_{555}$, $V_{606}$, or $I_{814}$
WFPC2 coverage over Abell 2390 and Abell 2667 reaches to only
$\sim26.6$ and $\sim26.5$ ($5\sigma$, 0.3''-diameter aperture) over
Abell 2390 and 2667, respectively.  These WFPC2 data are $\geq1$ mag
shallower than the ACS data available over 5 other clusters studied
here (Table~\ref{tab:obsdata}).  This will result in a much higher
contamination rate from low-redshift galaxies scattering into our
color selection due to noise.

To estimate the expected contamination level, we start with a
selection of intermediate magnitude galaxies in the HUDF NICMOS field
(Thompson et al.\ 2005) and add noise to the fluxes to match the
errors for sources in our search fields.  On average, we find $\sim1$
low-redshift contaminant over Abell 2390 and $\sim2$ such sources over
Abell 2667 in these simulations.  This is comparable to the observed
numbers and suggests that A2390-zD1, A2667-zD1, A2667-zD2, and
A2390-JD1 may be low-redshift interlopers.  As an alternate estimate
of the contamination level, we start with the catalog of sources in
the three clusters with the deepest ACS and NICMOS data (Abell 2218,
MS1358, CL0024) and add noise (again to match the flux errors for
sources behind Abell 2390 and Abell 2667).  On average, we predict
$\sim2$ low-redshift contaminants over Abell 2390 and $\sim3.5$
low-redshift contaminations over Abell 2667.  Again these numbers are
comparable to the observed number of candidates over these clusters
and suggest that all four candidates behind these clusters may be
interlopers.  The modest differences in the predicted number of
contaminants for the two methods is consistent with what one would
expect from small number statistics.

\textit{Surface Density of Robust $z\gtrsim7$ Candidates Behind
  Massive Galaxy Clusters:} We tabulate the surface density of robust
$z\gtrsim7$ candidates found behind clusters as a function of
magnitudes in Table~\ref{tab:surfdens}.  Also included in this table
are the surface densities found in the field (Bouwens et al.\ 2008;
R.J. Bouwens et al.\ 2008).  The surface density of $z\gtrsim7$
candidates behind clusters appears to be substantially larger than in
the field at bright magnitudes (i.e., $H<25.5$ or $>$5L$^*$).  This is
exactly what we expect as a result of the slope of the LF at bright
magnitudes.  Interestingly, current observations do not provide any
evidence for an enhancement in the surface density of $z\gtrsim7$
sources between $H\sim25.5$ and $H\sim27.0$.  This is despite the
expected steep slope of the LF at such magnitudes.  This could reflect
the small numbers of sources involved here and point to the need for
imaging more clusters to improve the overall statistics (see also \S6
and R.J. Bouwens et al.\ 2008, in prep).

At $H\gtrsim27$ AB mag we are probing faintward of $L^*$ and likely
reaching the regime where the effective slope of the LF is only
moderately steep (i.e., $-d(\log \Phi)/d(\log L)\sim0.7<1$: Bouwens et
al.\ 2007). For such slopes, the trade-off between depth and area is
such that the surface density of $z\gtrsim7$ dropouts behind clusters
will be lower than observed in the field.

\section{Discussion}

\subsection{Comparison with Model Expectations}

We can compare the number of $z\gtrsim7$ galaxy candidates found in
our search with that expected based on determinations of the $UV$
luminosity function (LF) at $z\sim7$ from the field (Bouwens et
al.\ 2008).  For this calculation, we need lensing models for the
clusters under study.  We have such lensing models for 5 of the search
clusters under study here, namely Abell 1689, Abell 1703, Abell 2218,
CL0024, MS1358.  These models are described in Limousin et
al.\ (2007), Limousin et al.\ (2008), El{\'{\i}}asd{\'o}ttir et
al.\ (2007), Jee et al.\ (2007), and Franx et al.\ (1997),
respectively.  We will scale our expectations from those clusters to
our entire sample.

\textit{Predicted Number of $z\gtrsim7$ Candidates:} Starting with the
Bouwens et al.\ (2008) model for the sizes, $UV$ colors, and $UV$ LF
for galaxies at $z\sim7$ (Appendix B from that work), we generate very
high-resolution mock images (pixel size $0.02''$) of galaxies in a
$z\sim7$ source plane and then remap these images to the image plane
using available lensing models for these clusters.  We then add these
simulated fields to the real data and attempt to recover these sources
using our cataloguing and selection procedures (\S3).  Generating
these mock fields 20 times for each cluster and using our selection
procedure to identify $z$ and $J$ dropouts, we estimate that we would
expect to find 2.7 $z$ dropouts over all of our search fields.  If we
repeat this procedure for $z\sim8-10$ $J$-dropouts assuming an
extrapolation of the Bouwens et al.\ (2008) LF results to $z\sim9$,
0.3 $J$ dropouts are expected.  We find only modest variations in the
number of dropouts expected behind clusters with similar depths and
survey areas (e.g., $\sim$0.2 predicted behind Abell 2218
vs. $\sim$0.3 predicted behind MS1358).  Similarly, use of slightly
different lensing models for the clusters only appears to have a
modest effect on the numbers (e.g., use of an updated version of the
Broadhurst et al.\ 2005 model only increases the predicted numbers
behind Abell 1689 by $\sim$20\% over that predicted by the Limousin et
al.\ 2007 model).

\textit{Comparisons against the Number of Observed $z\gtrsim7$
  Candidates:} The above predictions are quite consistent with the 1
robust $z$-dropout candidate, 3 other possible $z$-dropout candidates,
and 1 possible $J$-dropout candidate discussed in \S4, particularly
given the small numbers and uncertain contamination levels.  Because
of the difficulty in interpreting the search results over the clusters
with the shallower optical data (given the likely significant
contamination levels), it also makes sense to restrict our comparison
to only those clusters with the deepest data (i.e., excluding Abell
1835, Abell 2390, Abell 2667, and AC114 from this comparison).  Using
only clusters with the deepest optical data, we predict 2.0 $z\sim7$
$z$-dropouts and 0.2 $J$-dropouts in total.  Again this is in
reasonable agreement with the 1 robust $z$-dropout we find over this
more restricted search area, given the small number statistics.  For
both this comparison and the previous one, our search results are
clearly consistent with our predictions and therefore with the Bouwens
et al.\ (2008) $z\sim7$ LF derived in the field.  From this exercise,
it is also quite clear that $z\gtrsim7$ galaxy searches (using a small
number of clusters) provide only a very weak constraint on the volume
density of lower luminosity galaxies.  For our search the uncertainty
in the volume density of these sources is $\sim0.6$ dex [factor of
  four] based on our sample size of one.

\begin{deluxetable}{cccc}
\tablecolumns{4}
\tablewidth{3.4in}
\tablecaption{Surface density of strong $z\gtrsim7$ candidates in the deep near-IR data behind clusters and in the field.\label{tab:surfdens}}
\tablehead{\colhead{} & \colhead{} & \colhead{Surface} & \colhead{Search Area} \\
\colhead{} & \colhead{Magnitude} & \colhead{Density} & \colhead{per candidate} \\
\colhead{Location} & \colhead{Range\tablenotemark{a}} & \colhead{(arcmin$^{-2}$)} & \colhead{(arcmin$^2$)\tablenotemark{b}}}
\startdata
Cluster & $H_{160}<25.5$ & $0.05_{-0.04}^{+0.11}$\tablenotemark{c} & 21$_{-15}^{+101}$\\
Cluster & $25.5<H_{160}<26.5$ & $<$0.11\tablenotemark{d} & $>$9 \\
Cluster & $26.5<H_{160}<27.5$ & ---\tablenotemark{e} & --- \\
Field & $H_{160}<25.5$ & $<$0.03\tablenotemark{f} & $>$36 \\
Field & $25.5<H_{160}<26.5$ & $0.07_{-0.04}^{+0.07}$\tablenotemark{f} & 13$_{-6}^{+16}$ \\
Field & $26.5<H_{160}<27.5$ & $0.48_{-0.23}^{+0.38}$\tablenotemark{g} & 2$_{-1}^{+2}$ \\
\enddata
\tablenotetext{a}{Without any magnification from gravitational lensing and adopting the Bouwens et al.\ (2008) determination for $L^*$ at $z\sim7$, $H_{160}<25.5$ corresponds to luminosities $>5L_{z=7}^{*}$, $25.5<H_{160}<26.5$ corresponds to luminosities $>2L_{z=7}^{*}$ and $<5L_{z=7}^{*}$, and $26.5<H_{160}<27.5$ corresponds to luminosities $>0.8L_{z=7}^{*}$ and $<2L_{z=7}^{*}$.}
\tablenotetext{b}{This column is the reciprocal of the previous column.}
\tablenotetext{c}{Based on the NICMOS data over all 11 galaxy cluster fields considered in this paper}
\tablenotetext{d}{Based on the NICMOS data over Abell 1835, Abell 2218, Abell 2219, Abell 2390, Abell 2667, MS1358, CL0024, and AC114}
\tablenotetext{e}{Near-IR data over cluster fields are not deep enough
  to probe this regime well.  Nonetheless, we expect the surface
  density of $z\gtrsim7$ dropouts behind clusters to be lower than
  observed in the field, since at $H\gtrsim27$ AB mag we are probing
  faintward of $L^*$ and likely reaching the regime where the
  effective slope of the LF is only moderately steep (i.e., $-d(\log
  \Phi)/d(\log L)\sim0.7<1$: Bouwens et al.\ 2007).  For such a slope,
  the trade-off between depth and area is such that the surface
  density of dropouts is lower behind clusters.}
\tablenotetext{f}{Based on the Bouwens et al.\ (2008) and R.J. Bouwens
  et al.\ (2008, in prep) search results for $z\gtrsim7$ dropouts in the
  field.  The R.J. Bouwens et al.\ (2008, in prep) search takes
  advantage of more than 20 arcmin$^2$ additional search area with
  NICMOS not considered by Bouwens et al.\ (2008).}
\tablenotetext{g}{Based on data from the HUDF Thompson field (Thompson et
  al.\ 2005), HUDF-NICPAR1, and HUDF-NICPAR2 fields (Bouwens et
  al.\ 2008)}
\end{deluxetable}

\subsection{Comparison with Previous Results}

Richard et al.\ (2006) reported finding 13 $z\gtrsim6$ candidates
(1st+2nd category) in $\sim$12 arcmin$^2$ of ISAAC data behind 2
lensing clusters, while Richard et al.\ (2008) reported 12 other
$z\gtrsim7$ candidates in $\sim$9 arcmin$^2$ of NICMOS data behind six
other clusters.

The Richard et al.\ (2006) search results are significantly different
from our results (13 $z\gtrsim6$ candidates to $H_{160,AB}\sim25$ over
12 arcmin$^2$ vs. the one robust candidate we find over 21
arcmin$^2$).  As we discuss in detail in Appendix C of Bouwens et
al.\ (2008), $>90$\% of their $z\gtrsim6$ candidates appear to be
spurious, since none of the eleven $z\gtrsim6$ candidates in their
selection with substantially deeper ($\sim$1-2 mag) NICMOS+IRAC
coverage are significantly ($>$2$\sigma$) detected.

The prevalence of $z$$\gtrsim$7 galaxies implied by the Richard et al.
(2008) search results is also much greater than what we find in our
searches.  Since these candidates are reported over a subset of the
clusters used in the current search, the differences are puzzling.  To
understand the possible differences, we examined the Richard et
al. (2008) sources in our own reductions of the same HST data using
our photometric procedures.  We find that 2 of their 12 candidates
appear to be plausible $z\gtrsim7$ galaxies given our photometry.  The
other sources are likely contaminants (a detailed discussion of all 12
Richard et al. 2008 candidates can be found in Appendix A).  The two
good candidates also satisfy our $z$-dropout selection criteria (but
are blended with foreground galaxies in our search).  One of these two
$z\gtrsim7$ candidates does not have particularly deep optical
coverage, so there is a reasonable chance that it is at lower
redshift.

A comparison of our $z\gtrsim7$ sample with the Richard et al.\ (2008)
$z\gtrsim7$ sample suggests that our selection may suffer from some
incompleteness.  The observed level of incompleteness is not
surprising and is consistent with what we expect from blending with
foreground sources ($\sim$35\%).  Moreover, since this same
incompleteness is implicitly included in the predictions we make in
\S5.1 to compare with the observations, the conclusions that we draw
based upon those comparisons should be fair.\footnote{Even if we
  include the two best $z\gtrsim7$ candidates from the Richard et
  al.\ (2008) selection in our sample, the total number of strong
  $z\gtrsim7$ candidates is still very consistent with the predictions
  we make in \S5.1 from the Bouwens et al.\ (2008) field LFs.
  Moreover, for the clusters with the deepest optical data (i.e.,
  excluding AC114, Abell 1835, Abell 2390 and Abell 2667), only 2.2
  $z\gtrsim7$ candidates are predicted (\S5.1) which is quite
  consistent with the 2 found (i.e., A1689-zD1 and A2219-z1).}

\begin{deluxetable}{ccccc}
\tablecolumns{5}
\tablewidth{3.5in}
\tablecaption{Number of strong $z\gtrsim7$ candidates identified vs. the number of HST NICMOS orbits used for the search (see \S6).\label{tab:effic}}
\tablehead{\colhead{} & \colhead{} & \colhead{NICMOS} & \colhead{Candidates} & \colhead{Orbits per}\\
\colhead{Location} & \colhead{\#\tablenotemark{a}} & \colhead{Orbits} & \colhead{per Orbit} & \colhead{Candidate\tablenotemark{b}}}
\startdata
Cluster (Shallow\tablenotemark{c})\tablenotemark{e} & 1 & 144 & 0.007$_{-0.006}^{+0.016}$ & 144$_{-101}^{+682}$ \\
Cluster (Deep\tablenotemark{d})\tablenotemark{f} & 0 & --- & --- & --- \\
Field (Shallow\tablenotemark{c})\tablenotemark{g} & 3 & $\sim$450 & 0.007$_{-0.004}^{+0.007}$ & 150$_{-74}^{+178}$ \\
Field (Deep\tablenotemark{d})\tablenotemark{g} & 6 & $\sim$600 & 0.010$_{-0.004}^{+0.006}$ & 100$_{-37}^{+65}$ \\
Field (Both\tablenotemark{h}) & 9 & $\sim$1050 & $0.009_{-0.003}^{+0.004}$ & 117$_{-37}^{+56}$ \\
\enddata
\tablenotetext{a}{Number of strong $z\gtrsim7$ candidates}
\tablenotetext{b}{Reciprocal of the previous column}
\tablenotetext{c}{Search results for shallow near-IR data with $5\sigma$ depths less than $\sim$27 AB mag (i.e., $\leq5$ orbits).  27 AB mag corresponds to $\sim L^*$ at $z\sim7$ (Bouwens et al.\ 2008).}
\tablenotetext{d}{Search results for deep near-IR data, with $5\sigma$ depths greater than $\sim$27 AB mag (i.e., $\geq6$ orbits).  27 AB mag corresponds to $\sim L^*$ at $z\sim7$ (Bouwens et al.\ 2008).}
\tablenotetext{e}{All 11 clusters considered here.}
\tablenotetext{f}{No very deep (i.e., reaching $>$27 AB mag at $5\sigma$) near-IR data are available over clusters, but it is expected that the search efficiency will decrease significantly at fainter magnitudes since we will be probing faintward of $L^*$ where the faint-end slope is likely only moderately steep (\S6).}
\tablenotetext{g}{Based upon the NICMOS data considered in the Bouwens et al.\ (2008) and R.J. Bouwens et al.\ (2008, in prep) $z\gtrsim7$ searches}
\tablenotetext{h}{Search results for both shallow and deep near-IR data.}
\end{deluxetable}

\section{Implications}

The purpose of the present study is to increase the sample of
candidate $z$$\sim$7 galaxies and to assess the potential of
gravitational lensing by galaxy clusters to identify and quantify the
properties of galaxies at $z\gtrsim7$.  Lensing will increase the
depth of the survey by the magnification factors, but decrease the
search area by the same factor.  If the effective slope of the LF
($-d(\log d\Phi)/d\log L$) is greater than 1, the gain in depth more
than compensates for the loss in area, increasing the overall number
of sources (e.g., Broadhurst et al.\ 1995).  We would expect this
effect to increase the numbers at bright magnitudes where the
effective slope of the LF is very steep due to an apparent cut-off at
the bright end (i.e., $H\lesssim27$), but to decrease the numbers at
fainter magnitudes ($H\gtrsim27$) where the faint-end slope is
shallower than this (i.e., $-d(\log d\Phi)/d\log L\sim0.7\lesssim$1:
e.g., Bouwens et al.\ 2007).

Our simulations (\S5.1) show the expected gains at the bright end of
the LF for searches behind clusters (see Table~\ref{tab:effic}).  We
expect three $z\gtrsim7$ galaxies over the present set of cluster data
(144 NICMOS orbits) using the observed LF of Bouwens et al.\ (2008).
The 48 NICMOS orbits/galaxy in the clusters for bright sources
contrasts with the $\sim$120 NICMOS orbits/galaxy needed in the field
(nine $z\gtrsim7$ galaxies are found in $\sim$1050 orbits of NICMOS
data over the GOODS fields: R.J. Bouwens et al.\ 2008, in prep).
Nearly identical search procedures and selection criteria are used in
both the cluster and field searches.  Interestingly, the expected
gains at bright magnitudes in clusters are not reflected in the
observational results above, if we only include our most robust
candidate (for $\sim$144 NICMOS orbits/galaxy).  This could easily
arise because of small number statistics.  Any gains in using clusters
are likely to disappear at lower luminosities faintward of the LF
knee, since the slope is not steep enough, as noted above.

Of course, lensing clusters can be used to potentially detect objects
much fainter than in a field sample, thereby possibly extending the LF
function to fainter limits. Unfortunately, both small number
statistics and the challenges of modelling clusters make this very
difficult, and likely not to be a very practical approach.  This is
because a substantial sample of objects is needed to faint limits to
accurately determine the LF, as well as extremely accurate models of
both lensing by the foreground cluster and incompleteness suffered by
the lensed high-redshift population.  Determining either of these
latter two quantities well (and without any systematics) is a great
challenge. As a result, it can be difficult to even measure quantities
like the faint-end slope of the UV LF at $z$$\sim$4, $z$$\sim$5, and
$z$$\sim$6 from current samples of $g$, $r$, and $i$ dropouts behind
lensing clusters (where the samples are much larger: see R.J. Bouwens
et al.  2008, in prep).

Our findings also underline the importance of having very deep optical
data below the Lyman break for identifying high-z dropout galaxies.
Without such data, it is essentially impossible to distinguish
bona-fide $z\gtrsim7$ galaxies from the large number of low-redshift
galaxy interlopers that may scatter into $z\gtrsim7$ samples as a
result of noise -- as we found for the four weaker $z\gtrsim7$ dropout
candidates in our selection (\S4).  Unfortunately, obtaining
sufficiently deep optical data can be expensive and often requires
$\gtrsim2-3\times$ as much time as is spent obtaining the near-IR
data.

\acknowledgements

We thank Richard Ellis, Marusa Brada{\v c}, Johan Richard, Daniel
Schaerer, and Dan Stark for useful discussions.  We thank Johan
Richard for helping us to more precisely locate some of his team's
lower S/N $z$$\gtrsim$7 candidates in our own reductions of the
available NICMOS data.  We thank our referee for important feedback
that was valuable for improving the overall clarity of our manuscript.
This research used the facilities of the Canadian Astronomy Data
Centre operated by the National Research Council of Canada with the
support of the Canadian Space Agency.  We acknowledge support from
NASA grants HST-GO09803.05-A, HST-GO10874.04-A and NAG5-7697.

\appendix

\begin{figure}
\epsscale{1.01} \plottwo{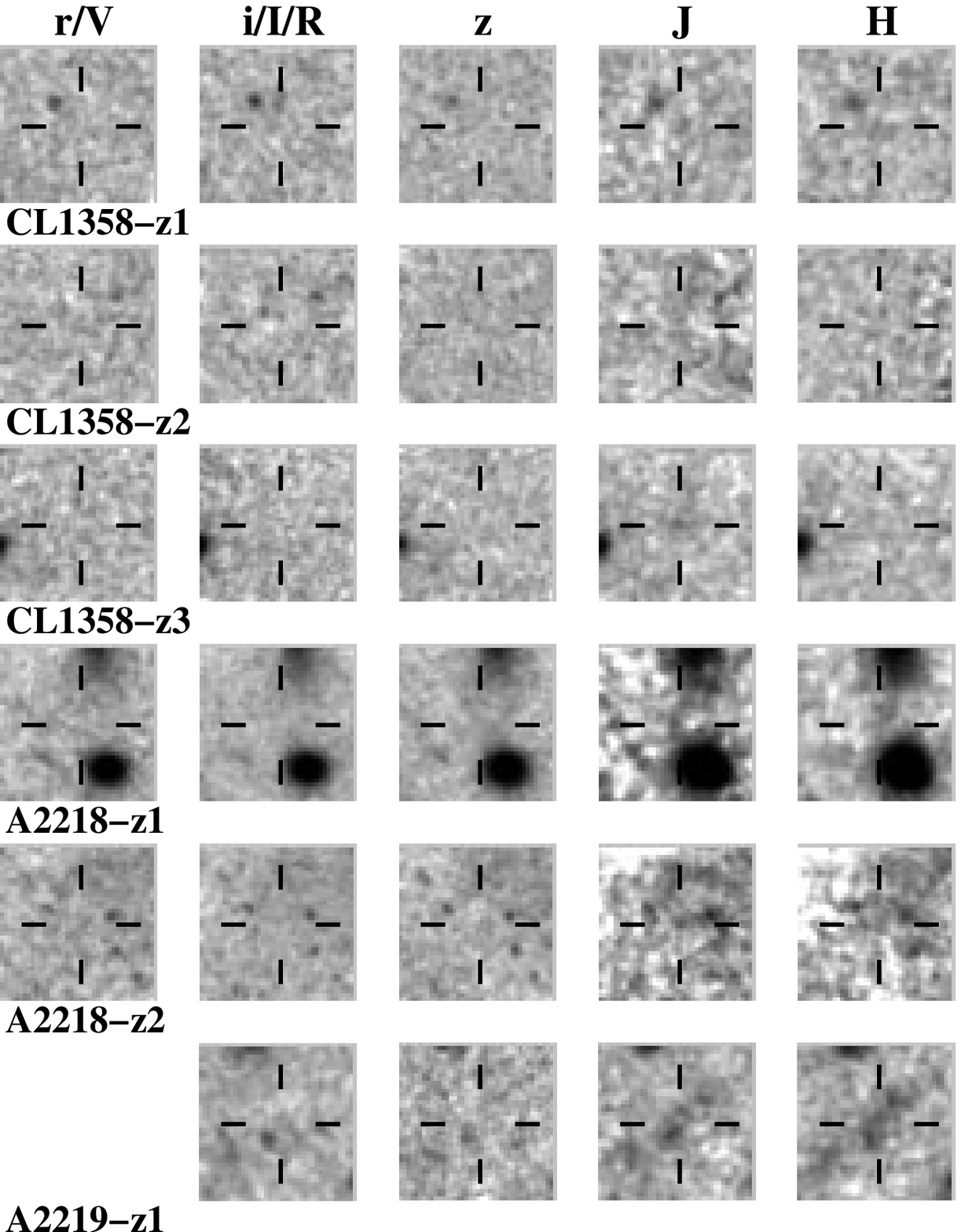}{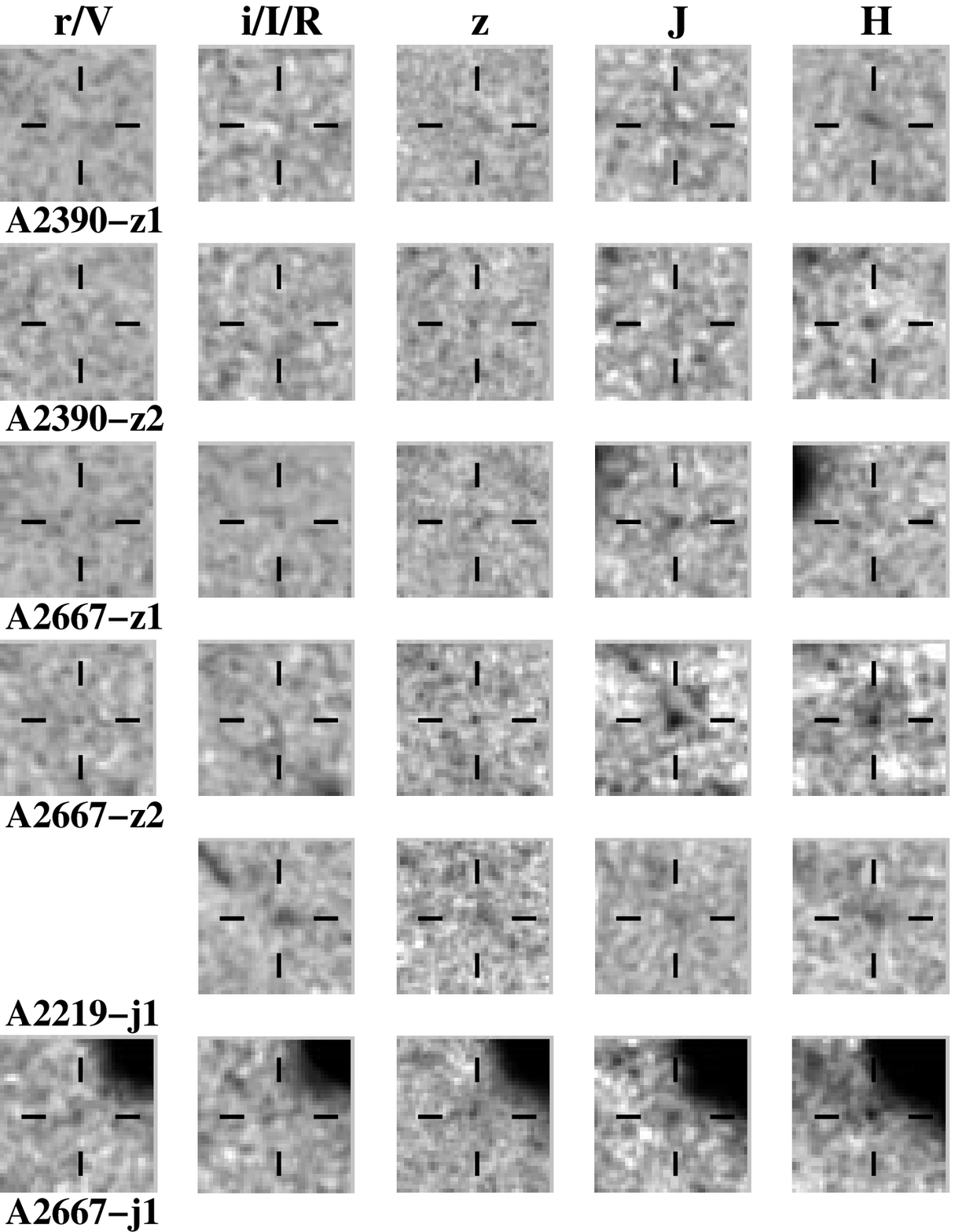}
\caption{$r_{625}$,$i_{775}$,$z_{850}$,$J_{110}$,$H_{160}$ images
  ($4'' \times 4''$) of the 12 $z$$\gtrsim$7 candidates in the Richard
  et al.\ (2008) selection.  The images are scaled for display in the
  same way as we used in Figure~\ref{fig:stamp} to present the
  $z\gtrsim7$ candidates we identified behind clusters (and as used in
  presenting $z\gtrsim7$ candidates in the HUDF/GOODS fields: Figure 3
  from Bouwens et al.\ 2008).  The leftmost two images correspond to
  the WFPC2 $V_{555}$ and $I_{814}$ bands for Abell 2390, the WFPC2
  $R_{702}$ band for Abell 2219, and the WFPC2 $V_{606}$ and $I_{814}$
  bands for Abell 2667.  Photometry for these candidates is presented
  in Table~\ref{tab:richardcand}.  A2219-z1 appears to be a blend of
  two sources, the upper right one of which appears to be a potential
  $z\gtrsim7$ candidate.  A2667-z2 satisfies our $z$-dropout selection
  criteria, but it is difficult to determine if it is a probable
  $z\gtrsim7$ galaxy given the limited depth of the optical data
  (Table~\ref{tab:obsdata}: see also \S4).  A2390-z1 and A2390-z2 have
  $z-J$ colours that appear to be too blue (see
  Table~\ref{tab:richardcand}) for them to be $z\sim7$ $z$-dropouts.
  The colors we measure for A2667-z1 do not satisfy our $z$-dropout
  selection criteria, but we cannot completely rule out this source as
  a $z\gtrsim7$ candidate.  CL1358-z1, CL1358-z2, and CL1358-z3 are
  not detected at $\geq3\sigma$ significance in our reductions of the
  NICMOS data (the possible sources seen in the $J$-band data have a
  significance of only $1-2.7\sigma$ in our reductions).  A2218-z1 is
  sufficiently close to a bright source, as to make unambiguous
  detection difficult.  A2218-z2 is only found after subtracting the
  wings of the central cluster galaxy and is detected at only modest
  significance ($\sim2-3\sigma$); the nature of this source is clearly
  very uncertain.  Both A2219-j1 and A2667-j1 show significant
  detections at optical wavelengths and are almost certainly low
  redshift contaminants.\label{fig:richardstamp}}
\end{figure}

\section{A.  Independent Assessment of the $z$$\gtrsim$7 candidates in Richard et al.\ (2008)}

Richard et al.\ (2008) also conducted a search for $z$$\gtrsim$7
galaxy candidates behind 6 of the 11 galaxy clusters considered in
this study (i.e., CL0024, CL1358, Abell 2218, Abell 2219, Abell 2390,
Abell 2667).  They reported finding 10 promising $z\sim7$ $z$-dropout
candidates and 2 promising $z\sim9$ $J$-dropout candidates.
Surprisingly, none of the 12 candidates in the Richard et al.\ (2008)
$z\gtrsim7$ sample make it into our own $z$$\gtrsim$7 dropout selections
(\S4) -- nor do our $z\gtrsim7$ candidates appear in their selection --
even though our color selection criteria are essentially identical
(and their search data a subset of ours).

To understand the reason for these differences, we performed
photometry on the sources using the same $0.3''$-diameter (for
ACS+WFPC2) and $0.6''$-diameter (for NICMOS) apertures employed by
Richard et al.\ (2008) for their photometry.  We then corrected the
magnitudes we measured from the ACS+WFPC2 or NICMOS data by 0.3 mag
and 0.6 mag, respectively (again following the methodology of Richard
et al.\ 2008).  For the four Richard et al. (2008) candidates where
light from the central cluster galaxy would contaminate the photometry
(i.e., A2218-z2, A2390-z2, A2667-z2, A2667-j1), we explicitly fit the
isophotes to the central galaxy and subtracted them off.  A summary of
our photometry of the 12 Richard et al.\ (2008) candidates is
presented in Table~\ref{tab:richardcand}.  We present images of the 12
$z$$\gtrsim$7 candidates from the Richard et al.\ (2008) selection in
Figure~\ref{fig:richardstamp}.

Of the 12 candidates, two (i.e., A2219-z1 and A2667-z2) seem like
plausible $z\gtrsim7$ galaxies (A2219-z1 being the stronger of the two).
Both candidates satisfy our selection criteria (but were blended with
foreground galaxies in our catalogs and therefore not included in our
candidate lists).  The other ten candidates appear unlikely to
correspond to high-redshift sources, given our photometry (although in
the case of A2667-z1 it is difficult to rule out the source being at
$z\gtrsim7$).  Ascertaining the precise redshift of A2667-z2 is difficult
given the limited depth of the optical data around Abell 2667.
Without deeper optical data, it is essentially impossible to know
which $z\gtrsim7$ candidates correspond to high-redshift galaxies and
which are lower redshift contaminants.  Note that we encountered
similar difficulties in ascertaining the nature of the 4 weaker
$z\gtrsim$7 candidates found in our selection behind Abell 2390 and
Abell 2667 (neither of which has particularly deep optical coverage).
Our simulations and tests (\S4) suggested that substantial
contamination by low redshift galaxies is quite likely.

In addition to the above comments about specific $z\gtrsim7$
candidates in the Richard et al.\ (2008) selection, we also have
several general concerns about the properties of this
$z_{850}$-dropout selection.  Our first concern regards the
$J_{110}-H_{160}$ color distribution of this selection.  The colors
are $\sim0.5$ mag bluer on average than what we find in the field for
our $z$-dropout sample (Bouwens et al.\ 2008: Figure A1).  They are
also $\sim0.5$ mag bluer than what one would expect for a young
$\beta=-2$ star-forming galaxy population (as observed at $z\sim5-6$:
e.g., Lehnert \& Bremer 2003; Stanway et al.\ 2005; Bouwens et
al.\ 2006).  It is hard to understand why $z$-dropouts from the
Richard et al. (2008) selection would be so much bluer than these
expectations unless the candidates had strong Ly$\alpha$ emission in
the $J_{110}$ band.  However, such an explanation would appear to be
ruled out by the follow-up spectroscopy that Richard et al.\ (2008)
conduct that find no such emission for the 7 $z$-dropout candidates
they observe.

Second, we find it worrisome that almost all (8 out of 10) of their
$z_{850}$-dropout candidates only satisfy their $z_{850}-J_{110}$
color criterion by a small margin ($<0.3$ mag) whereas in the Bouwens
et al.\ (2008) field sample, almost half of the $z$-dropouts satisfy
the $z_{850}-J_{110}$ color criterion by a wide margin (i.e., $>1$
mag).  One would expect the situation to be quite opposite here, as a
result of the much flatter number counts expected behind galaxy
clusters (the number of bright sources in cluster fields would
increase due to magnification by the cluster and the number of faint
sources would decrease because of the loss of area).  This should
result in a \textit{larger} fraction of sources satisfying the
selection criterion by a wide margin, not a smaller fraction.

\begin{deluxetable*}{cccccc}
\tablecolumns{6}
\tablecaption{Independent photometry on the Richard et al.\ (2008) $z\gtrsim7$ candidates.\label{tab:richardcand}}
\tablehead{
\colhead{} & \multicolumn{5}{c}{0.6$''$-diameter aperture\tablenotemark{a}}\\
\colhead{Object ID} & \colhead{$0.6\mu$$-$$J_{110}$\tablenotemark{b}} & \colhead{$0.8\mu$$-$$J_{110}$\tablenotemark{c}} & \colhead{$z_{850}$$-$$J_{110}$} & \colhead{$J_{110}$} & \colhead{$H_{160}$}}
\startdata
\multicolumn{6}{c}{$z\sim7$ $z$-dropouts} \\
CL1358-z1\tablenotemark{d} & $>$1.9 & $>$1.7 & $>$0.8 & 27.6$\pm$0.7 & 27.8$\pm$0.7 \\
CL1358-z2\tablenotemark{d} & --- & --- & --- & $>$28.0 & $>$28.1 \\
CL1358-z3\tablenotemark{d} & $>$2.2 & $>$2.0 & $>$1.2 & 27.2$\pm$0.4 & $>$28.3 \\
A2218-z1\tablenotemark{e,f} & 2.0$\pm$0.7 & $>$1.8 & $>$1.4 & 26.7$\pm$0.2 & 27.6$\pm$0.4 \\
A2218-z2\tablenotemark{e} & 0.9$\pm$0.7 & $>$0.9 & 0.5$\pm$1.0 & 27.4$\pm$0.5 & 26.9$\pm$0.4 \\
A2219-z1 (lower left)\tablenotemark{g} & --- & 1.6$\pm$0.2 & 0.5$\pm$0.3 & 26.0$\pm$0.2 & 26.0$\pm$0.2 \\
A2219-z1 (upper right)\tablenotemark{g} & --- & $>$3.0 & 1.0$\pm$0.5 & 26.4$\pm$0.3 & 26.3$\pm$0.2 \\
A2390-z1\tablenotemark{h} & $>$1.5 & $>$1.6 & 0.0$\pm$0.4 & 26.8$\pm$0.3 & 26.6$\pm$0.2 \\
A2390-z2\tablenotemark{h} & $>$1.4 & $>$1.3 & $-$0.4$\pm$0.7 & 27.4$\pm$0.6 & 26.7$\pm$0.3 \\
A2667-z1\tablenotemark{f,i} & $>$2.1 & $>$1.4 & 0.5$\pm$0.4 & 26.6$\pm$0.3 & 27.5$\pm$0.6 \\
A2667-z2\tablenotemark{i} & 2.1$\pm$0.7 & $>$2.0 & 0.9$\pm$0.3 & 25.6$\pm$0.1 & 25.9$\pm$0.1 \\
\multicolumn{6}{c}{$z\sim9$ $J$-dropouts} \\
A2219-j1\tablenotemark{j} & --- & 2.0$\pm$0.3\tablenotemark{k} & 1.2$\pm$0.5\tablenotemark{k} & 26.7$\pm$0.2 & 26.0$\pm$0.2 \\
A2667-j1\tablenotemark{j} & 0.8$\pm$0.4\tablenotemark{k} & $>$1.3\tablenotemark{k} & 0.4$\pm$0.4\tablenotemark{k} & 27.8$\pm$1.0 & 26.4$\pm$0.2 \\
\enddata
\tablenotetext{a}{Following the Richard et al.\ (2008) procedure, we have corrected the magnitudes we measure in a $0.3''$-diameter and $0.6''$-diameter aperture by 0.3 mag and 0.6 mag, respectively, for the WFPC2+ACS and NICMOS data (see Appendix A).}
\tablenotetext{b}{This colour corresponds to $r_{625}-J_{110}$ for CL1358 and Abell 2218, $V_{555}-J_{110}$ for Abell 2390, and $V_{606}-J_{110}$ for Abell 2667.}
\tablenotetext{c}{This colour corresponds to $i_{775}-J_{110}$ for CL1358 and Abell 2218, $R_{702}-J_{110}$ for Abell 2219, and $I_{814}-J_{110}$ for Abell 2390 and Abell 2667.}
\tablenotetext{d}{These sources are not detected at high significance
  (i.e., $>3\sigma$) in either the $J_{110}$ or $H_{160}$ bands in our
  NICMOS reductions.  One possible reason that our calculated
  significance levels may be different from Richard et al.\ (2008) is
  that we account for sensitivity variations across the NIC3 detector in 
  our weight maps (the sensitivities vary by factors of $\sim$2 
  from region to region).}
\tablenotetext{e}{These candidates lie close enough to bright sources, as to make unambiguous detection and robust measurement quite difficult.}
\tablenotetext{f}{The measured $J_{110}-H_{160}$ color here is much
  bluer than those found for $z\sim7$ galaxies in the field (Bouwens
  et al.\ 2008) -- strongly suggesting this source is not a $z\sim7$ galaxy.}
\tablenotetext{g}{A2219-z1 appears to be a blend of two sources (see Figure~\ref{fig:richardstamp}).  While the lower left source is clearly a low redshift source, the upper right source seems to be a plausible $z\gtrsim7$ candidate.}
\tablenotetext{h}{The $z-J$ colours we measure for these sources
  appear too blue to correspond to dropout galaxies at $z>7$.}
\tablenotetext{i}{These candidates have colours that are consistent with those of galaxies at $z\sim7$.  However, it is difficult to be sure given the limited depth of the optical data over Abell 2667 (Table~\ref{tab:obsdata}: \S4).}
\tablenotetext{j}{These sources are detected at $\geq$2$\sigma$ at optical wavelengths in our reductions, strongly suggesting that they do not correspond to star-forming galaxies at $z>7$.  Richard et al.\ (2008) also note this fact and concede that these two sources are not particularly compelling $z\gtrsim7$ candidates.}
\tablenotetext{k}{The colours tabulated here are relative to the $H$-band, not the $J_{110}$-band, and hence are $0.6\mu-H_{160}$, $0.8\mu-H_{160}$, and $z_{850}-H_{160}$.}
\end{deluxetable*}

\end{document}